# Self-seeded lasing in ionized air pumped by 800 nm femtosecond laser pulses

Yi Liu*, Yohann Brelet, Guillaume Point, Aurélien Houard,
and André Mysyrowicz[+]

*Laboratoire d'Optique Appliquée, ENSTA ParisTech/CNRS/Ecole Polytechnique, 828, Boulevard des Maréchaux, Palaiseau, F-91762, France*
\* yi.liu@ensta-paristech.fr;
\+ andre.mysyrowicz@ensta-paristech.fr

**Abstract**: We report on the lasing in air and pure nitrogen gas pumped by a single 800 nm femtosecond laser pulse. Depending on gas pressure, incident laser power and beam convergence, different lasing lines are observed in the forward direction with rapid change of their relative intensities. The lines are attributed to transitions between vibrational and rotational levels of the first negative band of the singly charged nitrogen molecule-ion. We show that self-seeding plays an important role in the observed intensity changes.

## 1. Introduction

Lasing in ambient air or its major components ($N_2$ and $O_2$) when pumped by an intense ultrashort laser pulse has attracted much attention in recent years [1-12]. The generation of coherent radiation with such cavity-less laser sources holds great potential for remote sensing applications. Q. Luo and coworkers were the first to suggest the existence of a *backward* stimulated emission from filaments formed in air by 800 nm femtosecond pulses, based on an exponential increase of nitrogen fluorescence with filament length [5]. Dogariu *et al*. demonstrated in 2011 that a backward coherent emission at 845 nm can be achieved in air pumped by picosecond UV pulses [1]. The stimulated emission was attributed to optical dissociation of $O_2$ molecules followed by multiphoton excitation of atomic lines of Oxygen. Shortly later, Kartashov *et al*. reported a strong backward lasing emission at 337 nm and 357 nm in pure Nitrogen gas and its mixture with Argon when driven by powerful mid-infrared lasers (λ = 3.9 μm or 1.03 μm), due to a collision-induced population inversion between the $C^3\Pi_u^+$ and $B^3\Pi_g^+$ states of the excited $N_2$ molecule [2-3]. Here, the electronic transitions of nitrogen molecule are the same as in the conventional discharge-pumped nitrogen laser [13].

In 2011, Yao *et al*. reported a *forward* stimulated radiation from $N_2$ molecule-ions pumped by short and intense mid-infrared pulses [7]. The lasing effect corresponded to the $B^2\Sigma_u^+ \to X^2\Sigma_g^+$ transition between the excited and ground cation states. It could be triggered by fine tuning of the wavelength of a femtosecond Optical Parametric Amplification (OPA) laser operating in the mid-IR (λ = 1.2 μm-2.9 μm). It was found that when the 3[rd] or 5[th] harmonic of the driving pulse overlapped with specific spectral positions of the above transitions, lasing radiation was initiated by a seeding effect. The authors attributed the lasing effect to the amplification of the weak harmonic pulse in the presence of population inversion [7, 9]. The mechanism responsible for population inversion is still a subject of controversy [7, 9, 11, 12]. Later, the same group showed that lasing action of $N_2^+$ could also be activated with 800 nm femtosecond laser pulses, but required injection of a seed pulse around 400 nm [11, 12]. This external 400 nm pulse is the counterpart of the 3[rd] and 5[th] harmonic in the case of mid-IR pumping. Lasing obtained with a pump at 800 nm is particularly interesting because of the wide availability of femtosecond lasers at this wavelength.

In this article, we report on forward lasing effect of $N_2^+$ in ambient air and pure nitrogen gas pumped by just one pulse at 800 nm, without any external seeding pulse. We attribute the observed lasing lines around 391 nm and 428 nm to a self-seeding effect. Lines around 391 nm and 428 nm have been attributed to $B^2\Sigma_g^+(v=0) \to X^2\Sigma_g^+(v'=0)$ and $B^2\Sigma_g^+(v=0) \to X^2\Sigma_g^+(v'=1)$ transitions. Specifically, the 391 nm transition is seeded by the 2$^{nd}$ harmonic around 400 nm generated by the laser pulses in the plasma. We attribute the 428 nm laser emission dominating at higher gas pressure to the seeding by the super-continuum generated when the intense incident laser pulse undergoes filamentation. It is further found that the gas pressure and incident laser beam convergence play an important role in determining the intensity of the emission lines. We thus confirm the importance of an injection source proposed by Yao *et al.* to explain the forward stimulated emission between cation states of $N_2$ and show in addition that several other factors such as gas pressure, nonlinear propagation effects, and the presence of oxygen play a key role for optical gain.

## 2. Experimental results and discussion

In the experiment, we employed a commercial Chirped-Pulse-Amplification laser system (Thales, Alpha 100), which provides 45 fs pulses at 800 nm with energy up to 12 mJ per pulse. The laser pulses were focused by a convex focal lens ($f$ = 40 cm, 75cm, or 150 cm) into a gas chamber with variable length. In the experiment, the chamber was first pumped to a pressure as low as 10$^{-5}$ mbar and then filled with air or nitrogen gas at different pressures. Visible plasma strings with length of 5 to 30 mm were observed in the gas chamber, with a string length depending on the focal lens, incident pulse energy, and gas pressure. The pulses emerging from the gas chamber after the plasma generation area were first filtered by two color filters (BG 40), which transmit the spectral components shorter than 600 nm but block largely the fundamental pulses and the accompanying super-continuum. The filtered pulses were then focused by a fused silica lens ($f$ = 10 cm) to couple them into a fibre spectrometer (Ocean Optics, HR 4000). A photodiode was also used to measure the radiation intensity. No significant emission was detected in the backward direction in our current experiment.

In Fig. 1 (a), we present the measured spectrum as a function of gas pressure in air in the case of $f$ = 40 cm. The incident laser pulse energy was 3.5 mJ. A strong narrow-line radiation around 391 nm is observed for a pressure ranging from 20 mbar to 200 mbar. The radiation around 391 nm has been identified previously as the transition between the vibration states $v$ = 0 of second excited state $B^2\Sigma_g^+$ of the nitrogen ions $N_2^+$ and its ground state $X^2\Sigma_g^+$ [7, 11, 12]. We find that the radiation around 391 nm is actually composed of two separated lines at 390 nm and 391.6 nm (Fig. 1 (b)). We attribute the peaks at 390 nm and 391.6 nm to the R-branch and P-branch of the $B^2\Sigma_g^+(v=0) \to X^2\Sigma_g^+(v'=0)$ transition, where different levels of quantum rotation of the molecular ions are involved [14, 15]. The relative strength of the two lines changes with pressure. At very low pressure, the P branch at 391.6 dominates, while at higher pressure $p$ > 150 mbar the R branch becomes comparable or even surpass the P branch.

We performed similar measurement in pure nitrogen gas. The results are presented in Fig. 1(c). Here the incident pulse energy was slightly higher, about 4 mJ. At the lowest pressure $p$ = 4 mbar, a weak broad emission peaked at 400 nm is observed, which will be discussed in more detail later. Under a slight increase of pressure, the lasing radiation around 391 nm appears, grows to a maximum for a pressure around 100 mbar and disappears above 200 mbar. Again, the same alternation between the relative importance of the P and R branch is observed with pressure. In contrast to air, the intensity of the lasing line around 428 nm increases significantly from 100 mbar up to $p$ = 750 mbar and then saturates for $p$ > 750 mbar. The line at 428 nm corresponds to the transition between the vibration states $v$ = 0 of second excited state $B^2\Sigma_g^+$ of the nitrogen ions $N_2^+$ and its $v$ = 1 ground state $X^2\Sigma_g^+$ [7, 11, 12]. A weak radiation peaked around 471 nm, corresponding to the $B^2\Sigma_g^+(v=0) \rightarrow X^2\Sigma_g^+(v'=2)$ transition [7, 9], is observed for pressure larger than 450 mbar. However, it does not grow into a strong emission for the pressure range tested. The evolution of the radiation intensity around 391 nm and 428 nm are presented in Fig. 2 (a) and (b) as a function of pressure.

We have repeated the same experiments in pure $N_2$ with longer focusing conditions. For focal distance $f$ = 75 cm, the intensity dependence of the two lasing lines as a function of pressure is shown in the Fig. 2 (c) and (d) for incident pulse energy of 3.5 mJ. The lasing emission around 391 nm appears, dominates, and disappears in the pressure range from 1 mbar to 50 mbar. At the same time, the lasing emission at 428 nm starts to show up at 15 mbar, reaches its maximum around 400 mbar, and then gradually decreases up to 1000 mbar. For $f$ = 150 cm no lasing emission was observed around 391 nm for all pressures tested while a relatively weak radiation around 428 nm was only observed for pressure below 80 mbar.

We now come back to the broad emission around 400 nm in the case of $p$ = 4 mbar, presented in Fig. 1(d). We attribute the observed broad radiation at 400 nm to the generation of second harmonic. In previous studies, 2$^{nd}$ harmonic generation in gas media was supposed to be impossible [11, 12], due to the isotropy of the medium. However, several groups reported second harmonic by simply focusing a femtosecond laser pulse in gases [16, 17]. The mechanism has been attributed to the nonuniform plasma distribution driven by the ponderomotive force or an electric-field-induced third-order mixing [16, 17]. To further check this hypothesis, we filled the chamber with low pressure $O_2$, Ar or $CO_2$ gases. For the three gases, a similar broad emission around 400 nm was observed for pressure below 80 mbar. The harmonic signal appears above an input pulse energy of 2 mJ. It rises with increasing pressure to reach a maximum around 15-30 mbar and then decreases progressively up to 80 mbar. A detailed study on the 2$^{nd}$ harmonic generation in gases will be published elsewhere.

Also, white-light continuum became visible to human eyes during the experiments for pressures larger than 100 mbar. Above 100 mbar, filamentation occurs under our experimental conditions. Broadband continuum generation accompanying the formation of filaments is a well known phenomenon. The presence of the supercontinuum is only weakly observed in Fig. 1 (a) around 490 nm for $p$ > 600 mbar, because the supercontinuum was strongly attenuated by the two BG40 filters. Without filters, we observed that the supercontinuum extends well below 400 nm in the case of $p$ = 1000 mbar both in air and pure $N_2$.

We measured the lasing radiation intensity in $N_2$ as a function of incident laser pulse energy for the two lasing lines. The results are presented in Fig. 3. The chosen nitrogen gas pressures for the two experiments were 100 mbar and 600 mbar for 391 nm and 428 nm, in view of their pressure dependence (Fig. 1 (c)). With the current detection system, incident laser pulse energy thresholds around 2 mJ were observed for both lasing lines. This corresponds to the threshold for appearance of the second harmonic.

A discussion of this behavior requires proper inclusion of nonlinear propagation effects. Due to nonlinear propagation, there is no simple relation between the incident pulse energy and the local laser intensity in the interaction region where plasma is produced. For instance, at the lowest pressures, the intensity at the focus is fixed by diffraction of the beam, because beam defocusing effects are negligible. The peak incident laser intensity then scales like the incident pulse energy and reaches a maximum value of $7.5 \times 10^{15} W/cm^2$ for a pulse energy of 2 mJ. At such intensities, a large fraction of the molecules are singly ionized. At higher pressures, defocusing of the beam limits the peak intensity. An increase of pulse energy does not change significantly the fraction of ionized molecules but extends the range of ionization (onset of filamentation). In filaments, the peak laser intensity is clamped to a value $I \sim 3 \times 10^{13}$ W/cm$^2$ for air or $N_2$ gas pressure between 200 and 1000 mbar [18].

We now try to explain the evolution of the observed self-seeded lasing effects presented in Fig. 1 (c) and compare it with the previous results of other groups. In the previous report of Yao and coworkers, where the incident pulse energy was below the energy threshold (~ 2 mJ) for harmonic generation, population inversion between the $B^2\Sigma_g^+$ state and the $X^2\Sigma_g^+$ state could be achieved [11, 12]. However, external seeding pulse was required to initiate lasing phenomenon [11, 12]. In our case, the incident laser energy exceeds the threshold energy for $2^{nd}$ harmonic generation. As a result, effective $2^{nd}$ harmonic around 400 nm is generated (Fig. 1(d)), which serves as the seeding pulse to start the lasing around 391 nm. With pressure beyond 100 mbar, the 391 nm lasing radiation decreases due to the gradual yield decrease of the $2^{nd}$ harmonic. At the same time, the lasing activity at 428 nm takes over, thanks to the effective generation of supercontinuum, which gradually approaches 428 nm with increasing pressure. At still higher pressures, for example, $p > 750$ mbar in Fig. 1 (c) and Fig. 2 (b), and $p > 400$ mbar in Fig. 2 (d), the lasing intensity saturates or even decreases. At higher pressure, the supercontinuum seed is expected to be stronger. Therefore, this suggests that the optical gain, *i.e.* the population inversion between the two involved states, decreases at higher pressure.

Finally, we measured the spatial profile of the lasing radiation as a function of nitrogen gas pressure. Results are presented in Fig. 4 in the case of $f = 75$ cm. In the experiment, a Charge-Coupled-Digital (CCD) chip was installed closely behind the two BG 40 filters to record the spatial profile of the transmitted lasing emissions. In Fig. 4 (a)-(c), the profiles of the lasing radiation around 391nm are shown for gas pressure of 5 mbar, 15 mbar, and 25 mbar. At the lowest pressure of 5 mbar, a Gaussian radiation pattern is observed (Fig. 4 (a)). Upon increase of pressure, an outer ring structure surrounding the central spot appears. The outer ring is better seen in the line-cuts shown in Fig. 4 (b) and (c). For the radiation peaked at 428 nm, a ring structure is observed at all pressures between 100 mbar and 1000 mbar (Fig. 4 (d)-(f)). The divergence angle of the lasing radiation was measured to be ~ 20 mrad in Fig. 4 (e). At this moment, we speculate this ring-shaped profile around 428 nm reflects the spatial distribution of the super-continuum seeding pulse, which is emitted in a conical form during laser pulse filamentation [19, 20].

## 3. Conclusion

In conclusion, we demonstrated that lasing radiation of $N_2^+$ can be achieved in ambient air or pure nitrogen gas pumped with a single 800 nm femtosecond laser pulse, without a second external seeding pulse. We attribute the initiation of the lasing effect to self-generated seeding pulses, either the $2^{nd}$ harmonic or the super-continuum white light. Gas pressure was shown to have a significant role on the lasing intensity and the spatial profile. The discovery of the self-seeded regime of the air plasma lasing pumped by the widely available 800 nm femtosecond pulses simplifies the experimental setup required and may prove useful for applications.

**Acknowledgments** The authors are grateful to Hongbing Jiang of Peking University, Benjamin Forestier of CILAS, Jinping Yao of SIOM, Thierry Lefrou of LOA for stimulating discussion and technical help. The project has been partially funded by Contract No. ANR-2010-JCJC-0401-01.

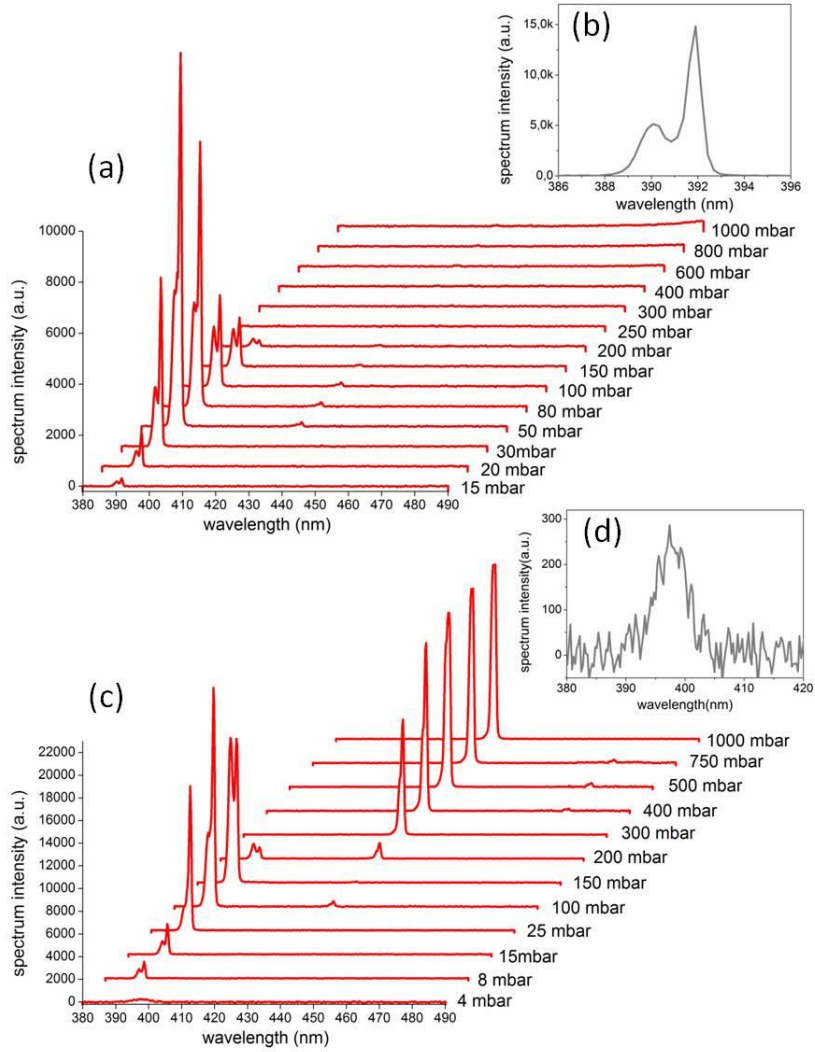

Fig. 1. (a) and (c), Spectrum intensity of the forward lasing radiations in air (a) and nitrogen gas (c) as a function of gas pressure. The focal length was $f = 40$ cm. The incident laser energy was 3.5 mJ in (a) and slightly higher, about 4 mJ, in (c). (b) Zoomed spectrum around 391 nm for pressure $p = 30$ mbar in Fig. 1(a). (d), Zoomed spectrum around 400 nm for pressure $p = 4$ mbar in Fig. 1 (c).

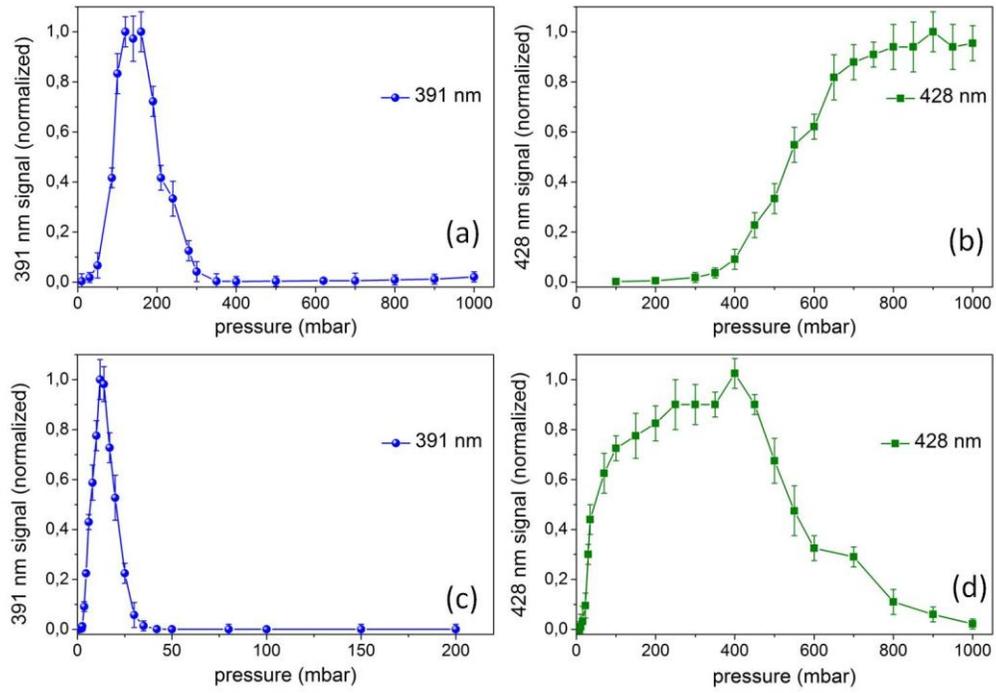

Fig. 2. Intensity of the lasing radiation around 391 nm (left) and 428 nm (right) as a function of nitrogen gas pressure. The focal length was 40 cm in (a) and (b), 75 cm in (c) and (d).

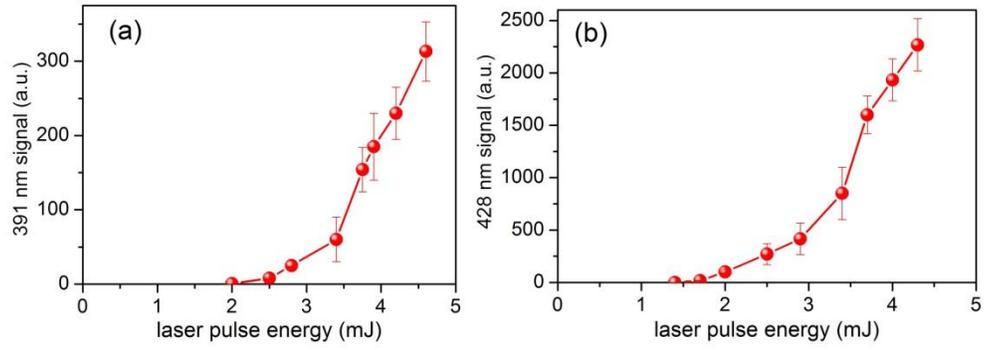

Fig. 3. Intensity dependence of the 391 nm (a) and 428 nm (b) lasing radiation on the incident laser pulse energy. The nitrogen pressure for (a) and (b) are 150 mbar and 600 mbar, respectively.

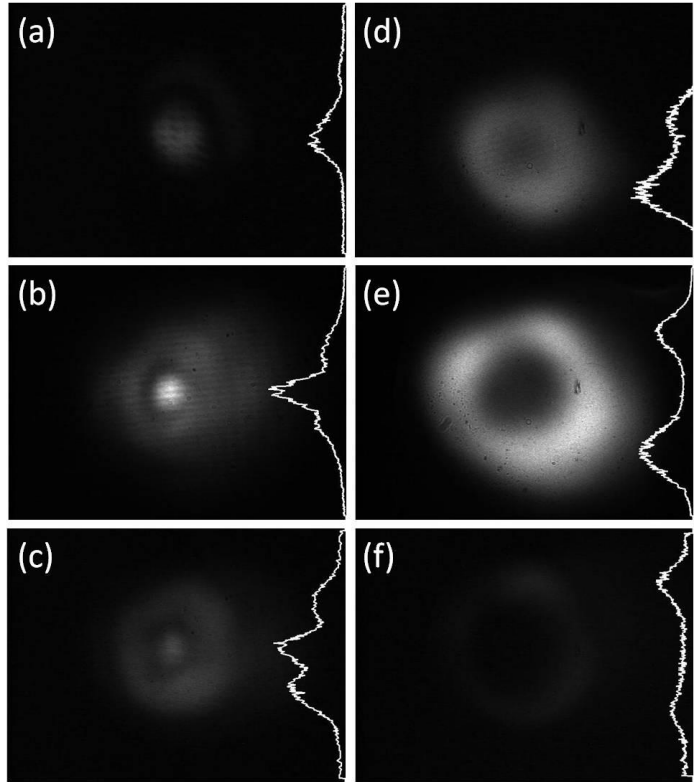

Fig. 4. (a)- (c), Spatial profile of the 391 nm emission at different nitrogen pressure. From (a) to (c), the pressure are 5 mbar, 15 mbar, and 25 mbar, respectively. (d) – (f), Spatial profile of the 428 nm lasing radiation. The pressures are 100 mbar, 300 mbar, and 800 mbar from (d) to (f). The angle of view of each panel is 60 × 40 mrad. The white lines on the right of each panel present the vertical line-cut of each intensity distribution on the center.